\documentclass[conference]{IEEEtran}

\IEEEoverridecommandlockouts
\usepackage{cite}
\usepackage{amsmath,amssymb,amsfonts,comment}
\usepackage{algorithmic}
\usepackage{graphicx}
\usepackage{textcomp}
\usepackage{xcolor,balance}
\usepackage[ruled,vlined]{algorithm2e}
\def\BibTeX{{\rm B\kern-.05em{\sc i\kern-.025em b}\kern-.08em
  T\kern-.1667em\lower.7ex\hbox{E}\kern-.125emX}}

\newcommand{\tr}[1]{\mathrm{tr}\!\left[{#1}\right]}

\begin{document}

\title{An IRS-Assisted Secure Dual-Function Radar-Communication System  
}

\author{
	\IEEEauthorblockN{Yi-Kai Li$^{1,2}$ and Athina Petropulu$^1$}
	\IEEEauthorblockA{$^1$Dept. of Electrical and Computer Engineering, Rutgers University, Piscataway, NJ, USA\\
 $^2$Dept. of Electrical and Computer Engineering and Technology, Minnesota State University Mankato, MN, USA
		\\
		E-mail: \{yikai.li, athinap\}@rutgers.edu\vspace{-0mm}}
		\thanks{This work is supported by ARO grants W911NF2110071 and
W911NF2320103, and by  
NSF under grant ECCS-2033433.}
		}

\maketitle

\begin{abstract}
In dual-function radar-communication  (DFRC) systems the probing signal contains information intended for the communication users, which makes that information vulnerable to eavesdropping by the targets.  We propose a novel design for enhancing the physical layer security (PLS) of DFRC systems, via the help of intelligent reflecting surface (IRS) and artificial noise (AN), transmitted along with the probing waveform. The radar waveform, the AN jamming noise and the IRS parameters are designed to  optimize the communication secrecy rate  while meeting radar signal-to-noise ratio (SNR) constrains. 
\textcolor{black}{Key challenges in the resulting optimization problem include  the fractional form objective, the  SNR being  a quartic function of the IRS parameters, and the unit-modulus constraint of the IRS parameters. 
A fractional programming technique is used to transform the fractional form objective of the optimization problem into more tractable non-fractional polynomials.}
Numerical results are provided to demonstrate the convergence of the proposed system design algorithm, and also show the impact of the power assigned to the AN 
 on the secrecy performance of the designed  system.

\end{abstract}


\begin{IEEEkeywords}
DFRC, physical layer security, IRS
\end{IEEEkeywords}

\section{INTRODUCTION}
{
Integrated sensing and communication (ISC) systems aim to perform both sensing and communication functions from a common platform \cite{Chen2020performance,Feng2020joint,Zhang2021anoverview,Sturm2011waveform}. Thus, ISC systems are prime candidates for next-generation wireless systems, such as unmanned aerial vehicles or autonomous vehicles, which are envisioned to achieve both high data rates and high sensing performance simultaneously.
A special case of ISC systems is the Dual-Function Radar-Communication (DFRC) systems, which not only use a common platform for both communication and sensing but also use a shared waveform for these two functions \cite{Liu2020joint,Hassanien2019dual,Ma2020joint}.}
%
In DFRC systems, user information is embedded in the probing waveform, resulting in higher spectral efficiency compared to general ISC systems. However, this also raises security concerns as the embedded communication information can potentially be intercepted by a target that is also an eavesdropper. This paper focuses on  security issues associated with DFRC systems, approaching the design of the system from a physical layer security (PLS) perspective.

{By exploiting the physical characteristics of the wireless channel, PLS design aims to enable the legitimate destination to obtain the source information successfully, while preventing an eavesdropper (ED) from decoding the information {\cite{Wyner1975thewire}}.
PLS design of  communication systems has been well investigated \textcolor{black}{\cite{Fakoorian2011solutions,Dong2010improving,Zheng2011optimal,Li2011oncooperative,Zheng2013improving,Khisti2010secure,Goel2008guaranteeing}. One approach to ensure PLS is  {cooperative jamming}, where 
trusted relays act  as helpers and  beamform artificial noise (AN), aiming to degrade the ED's channel \cite{Fakoorian2011solutions,Dong2010improving,Zheng2011optimal,Li2011oncooperative,Zheng2013improving}. Another approach is that the source beamforms AN along with the information for users, in a way that the users do not experience interference \cite{Khisti2010secure,Goel2008guaranteeing}.
}
PLS design for DFRC systems has been considered in \cite{Su2021secure,Su2023sensing}, 
 where the DFRC system  embeds AN in the transmit waveform, and a radar beamformer and the AN are jointly designed respectively to minimize the ED signal-to-noise-ratio (SNR) \cite{Su2021secure}, and to maximize the weighted sum of normalized fisher information matrix determinant and normalized secrecy rate \cite{Su2023sensing}.

Here, we consider the problem of DFRC systems aided by intelligent reflecting surfaces (IRS). 
IRS can assist DFRC systems in overcoming performance limitations caused by path loss or blockage, which are likely to arise in the next-generation wireless systems. These systems often rely on high frequencies to achieve high communication rates and sensing performance. However, high frequencies experience high attenuation and blockage. 
 IRS  is a planar array that consists of passive elements, each of which can alter the phase of the incoming electromagnetic wave in a computer-controlled manner. These elements can cooperatively perform beamforming to increase the power level in intended directions or decrease it in unintended directions. 
{IRS can also create additional links between the radar and the users, or between the radar and the targets, thus introducing  more degrees of freedom (DoFs) for  system design \cite{Wei2022multiple,Jiang2021intelligent,Li2022dual,Li2023minorization,Liu2022joint}. When there is no   line-of-sight (LOS) between the radar and the target, the IRS can provide alternative paths for the radar signal to reach the target \cite{Wei2022multiple,Li2022dual}.}  

\textcolor{black}{In this paper, we investigate design for an IRS-aided secure DFRC system design from the PLS perspective. The radar transmits a precoded waveform along with additive precoded AN. The  precoding matrices and the IRS parameter matrix are jointly designed in order to optimize
 the communication secrecy rate, reflecting secure communication performance, while maintaining a certain level of  radar SNR, reflecting good target sensing performance.}
The secrecy rate is a non-convex  function of ratios 
and the radar SNR is  a non-convex fourth order function of the IRS parameter. 
\textcolor{black}{For a fixed IRS parameter, the design of waveform and AN precoding matrices can be formulated as a quadratic programming problem.}
\textit{The challenge  mainly lies in optimizing with respect to the IRS parameter, i.e., in simultaneously considering the non-convex fractional objective of secrecy rate, and the non-convex high-order  radar SNR.}
\textcolor{black}{
To address the non-convex high-order radar SNR term we  express it as a quadratic function  of an auxiliary variable, which is quadratic in the IRS parameter.  Subsequently,  we replace the SNR with a lower bound  that is linear in the auxiliary variable, and which is found via the  minorization technique \cite{Sun2017majorization}.
To address the non-convex fractional objective of secrecy rate and achieve a  tractable design problem we  invoke a fractional programming technique \cite{Shen2018fractional}.}}

 The literature on striking a balance between PLS and sensing performance for IRS-aided DFRC systems is sparse. In \cite{Salem2022active,Chen2022intelligent,Mishra2022optm3sec}, \textcolor{black}{the waveform design problems are formulated as quadratic programming problems, and respectively addressed by fractional programming technique \cite{Salem2022active}, Riemannian conjugate gradient (RCG) algorithm \cite{Chen2022intelligent}, and first-order Taylor series approximation \cite{Mishra2022optm3sec}. The IRS parameter design problems are respectively solved by minorization maximization (MM) \cite{Salem2022active}, the RCG algorithm \cite{Chen2022intelligent}, and the numerical simultaneous perturbation
stochastic approximation method \cite{Mishra2022optm3sec}.
}
\textcolor{black}{As compared to the above literature, our work addresses the  challenging fourth order SNR term, which is 
either not considered in \cite{Salem2022active,Chen2022intelligent}, or it is not  convexified to facilitate a tractable problem in \cite{Mishra2022optm3sec}.
As compared to \cite{Salem2022active,Chen2022intelligent}, besides the radar waveform design, we consider the design of AN as well, which results in a different design problem.
}

\vspace{-0mm}
\section{SYSTEM MODEL}
\label{sec:models}

\begin{figure}[!t]\vspace{0mm} 
	\hspace{6mm} 
	\def\svgwidth{240pt} 
	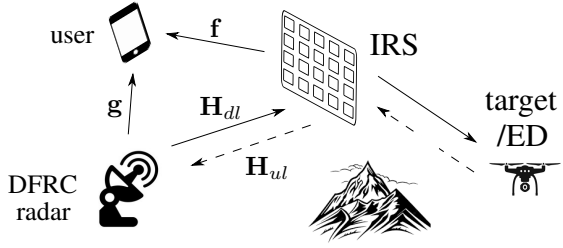 \vspace{-8mm}
	\caption{IRS-assisted secure DFRC system.}\vspace{-5mm} 
	\label{fig:system_model}
\end{figure}

We consider an IRS-aided DFRC system as Fig. \ref{fig:system_model}, where the DFRC is communicating with a user and is simultaneously tracking a non-line-of-sight (NLOS) target, who is also an eavesdropper. 
The DFRC has an $N_T$-antenna uniform linear array (ULA) transmitter, and a collocated $N_R$-antenna ULA receiver. The inter-antenna distance in both arrays is denoted by  $d$.   An $N$-element IRS is deployed to aid both the sensing and communication functionalities. The channels are assumed flat fading and perfectly known.
 The transmitted signal at the DFRC radar is
\begin{eqnarray}
	\mathbf x = \mathbf w  s + \mathbf W_n  \mathbf n,
\end{eqnarray}
\noindent where $\mathbf w \in \mathbb C^{N_T \times 1}$ denotes the   precoder for  information intended for the communication user, $s$ represents the transmit waveform that contains the information for the user. $s$ is assumed to be zero-mean white with unit variance; $\mathbf W_n \in \mathbb C^{N_T \times N_T}$ is the precoding matrix for the AN, and $\mathbf n \sim \mathcal{CN}(\mathbf 0_{N_T \times 1}, \sigma^2 \mathbf I_{N_T})$ is the AN. The AN is used for jamming and detecting the target/ED.

 Since there is no LOS between the radar and the target,  the transmitted waveform arrives at the target via the DFRC-IRS-target path, and the target echo reaches the radar receiver via the target-IRS-DFRC path. The signal at the radar receiver is

\vspace{-3mm}
\begin{small}
\begin{eqnarray} 
   \mathbf y_R &=& 
    {\beta \mathbf H_{ul} \mathbf \Phi \mathbf a_I(\psi_{a},\psi_{e}) \mathbf a_I^{T}(\psi_{a},\psi_{e}) \mathbf \Phi \mathbf  H_{dl} }
    (\mathbf w  s + \mathbf W_n  \mathbf n)+ \mathbf n_R \nonumber\\  &=& \mathbf C_{T} (\mathbf w  s + \mathbf W_n  \mathbf n) + \mathbf n_R,  \label{eqn:y_R}
\end{eqnarray}
\end{small}
\vspace{-3mm}

\noindent where $\beta$ is \textcolor{black}{the target reflection coefficient}; $\mathbf H_{ul}$ and $\mathbf H_{dl}$ are the normalized IRS-DFRC and DFRC-IRS channels, respectively; $\mathbf \Phi = \text{diag}([e^{j \phi_1},e^{j \phi_2},\cdots,e^{j \phi_N}])$ represents the IRS parameter matrix and is diagonal, where $\phi_n \in [0,2 \pi)$ is the phase shift of the $n$-th element of IRS; $\mathbf a_I(\psi_a,\psi_e)$ is the steering vector of IRS, and $\psi_a$ and $\psi_e$ are the azimuth and elevation angles of the target relative to the IRS, respectively; $\mathbf n_R \sim \mathcal{CN}(\mathbf 0_{N_R \times 1}, \sigma_R^2 \mathbf I_{N_R})$ indicates the additive white Gaussian noise (AWGN) at the DFRC receiving array, and $\sigma_R^2$  represents the   power of noise per radar receive antenna. 

The SNR at the radar receiver for detecting the target is
\begin{eqnarray}
	\gamma_R= \tr{ \mathbf C_{T} \left[ \mathbf w \mathbf w^H + \mathbf W_n \mathbf W_n^H \right] \mathbf C_{T}^H }/ \sigma_R^2. \label{eqn:gamma_R}
\end{eqnarray}
The received signal at the communication user is written as

\vspace{-3mm}
\begin{eqnarray} \label{eqn:y_U}
	 y_{u} &=& (\mathbf g^T + \beta_H^{\frac{1}{2}} \mathbf f^T \mathbf \Phi \mathbf H_{dl} ) (\mathbf w  s + \mathbf W_n  \mathbf n) +  n_u \nonumber \\ &=& \mathbf c_u^T (\mathbf w  s + \mathbf W_n  \mathbf n) +  n_u,
\end{eqnarray}
\vspace{-3mm}

\noindent where $\mathbf g \in \mathbb C^{N_T \times 1}$ is the  channel between the user and DFRC transmitter; $\beta_H$ is the path-loss of the DFRC-IRS channel; $\mathbf f \in \mathbb C^{N \times 1}$ denotes the   user-IRS channel; $n_u \sim \mathcal {CN}(0,\sigma_u^2)$ is   AWGN  at the user.

The received signal at the target/ED is

\vspace{-3mm}
\begin{eqnarray} \label{eqn:y_TE}
	y_{te} &=& ( \beta^{\frac{1}{2}} \mathbf a_I^T(\psi_{a},\psi_{e}) \mathbf \Phi \mathbf H_{dl} ) (\mathbf w  s + \mathbf W_n  \mathbf n) +  n_{te} \nonumber \\ &=& \mathbf c_{te}^T (\mathbf w  s + \mathbf W_n  \mathbf n) +  n_{te},
\end{eqnarray}
\vspace{-3mm} 

\noindent where $n_{te} \sim \mathcal {CN}(0,\sigma_{te}^2)$ is   AWGN  at the target/ED.

\noindent The achievable rate at the user and target/ED are respectively
\begin{eqnarray}
\!\!\!\!\!\!\!\!\!\!\!\!\!\!\!	&&R_u = \log \left( 1+\text{SINR}_{u} \right) = \log \left( 1+\frac{| \mathbf c_u^T \mathbf w|^2 }{ || \mathbf c_u^T \mathbf W_n||^2 + \sigma_u^2 } \right), \label{eqn:R_u} \\ \!\!\!\!\!\!\!\!\!\!\!\!\!\!\!&& R_{te} = \log \left( 1+\text{SINR}_{te} \right) = \log \left( 1+\frac{| \mathbf c_{te}^T \mathbf w|^2 }{ || \mathbf c_{te}^T \mathbf W_n||^2 + \sigma_{te}^2 } \right). \label{eqn:R_te}
\end{eqnarray}

\section{SYSTEM DESIGN}
\label{sec:sys_design}
We consider the  design of the radar information precoder, $\mathbf w$, the artificial noise precoder, $\mathbf W_n$, and the IRS parameter matrix, $\mathbf \Phi$,
as that of maximizing the secrecy rate,  defined as $(R_u-R_{te})$, while satisfying certain constraints, i.e.,
\begin{subequations} \label{eqn:orig_prob}
	\vspace{-0.1in}
	\begin{eqnarray} 
		\!\!\!\!\!(\mathbb P) \;\;\;\;\;\; \max_{\mathbf w, \mathbf W_n, \mathbf \Phi}&&	\;\; R_u - R_{te} \label{eqn:obj_sr}  \\ \!\!\!\!\!\mathrm{s.t.} && \;\; \tr{\mathbf w \mathbf w^H+\mathbf W_n \mathbf W_n^H}  \leq P_R \label{eqn:tot_power}\\\!\!\!\!\!&& \;\;  |\mathbf \Phi_{n,n}| = 1,\;\; \forall n = 1, \cdots, N\label{eqn:unit_modu}\\\!\!\!\!\!&& \;\;  \gamma_R  \geq \gamma_{R,th} \label{3d}    
		\label{eqn:radar_snr}
	\end{eqnarray} 
\end{subequations}

\noindent where (\ref{eqn:tot_power}) enforces that the total power of the waveform and AN stays below $P_R$; (\ref{eqn:unit_modu}) enforces unit modules  elements in $\mathbf \Phi$. This is because the IRS is composed of passive elements, which can only change the phase of the impinging signal; (\ref{eqn:radar_snr}) ensures that the radar SNR will  be above threshold $\gamma_{R,th}$.

The problem $(\mathbb P)$   in (\ref{eqn:orig_prob}) is challenging for the following reasons: (i) The design variables,  $\mathbf w, \mathbf W_n, \mathbf \Phi$, are mutually coupled. (ii) The objective is the difference between two non-convex functions ($R_u$ and $R_{te}$), and is non-convex. (iii) The radar SNR in  (\ref{eqn:radar_snr}), $\gamma_R$, is quartic in $\mathbf \Phi$. Since $\gamma_R$ is quadratic in $\mathbf C_{T}$ as (\ref{eqn:gamma_R}), and $\mathbf C_{T}$ is second order in $\mathbf \Phi$  as (\ref{eqn:y_R}). (iv) $\mathbf \Phi$ is subject to highly non-convex unit modulus constraints (UMC) as  (\ref{eqn:unit_modu}). 
    
To decouple the design variables, we divide the problem (\ref{eqn:orig_prob}) into two sub-problems. The first one is optimization with respect to  $\mathbf w$ and $\mathbf W_n$ for fixed  $\mathbf \Phi$, and the second one is optimizations with respect to $\mathbf \Phi$ for fixed  $\mathbf w$ and $\mathbf W_n$. These two sub-problems are solved in an alternating way until a convergence condition is met \cite{Li2017_coexistence}.

\subsection{First sub-problem: Solve for $\mathbf w$ and $\mathbf W_n$ by fixing $\mathbf \Phi$}

The first sub-problem is formulated as follows:

\vspace{-5mm}
\begin{subequations} \label{eqn:problem1}
	\begin{eqnarray} 
	\!\!\!\!\!\!\!\!\!(\mathbb P_{1})	\qquad \max_{\mathbf w, \mathbf W_n }&&	\;\;\;\;	R_u - R_{te}\label{eqn:obj1}\\ \!\!\!\!\!\!\!\!\!\mathrm{s.t.} && \;\;\;\; \tr{\mathbf w \mathbf w^H + \mathbf W_n \mathbf W_n^H}  \leq P_R. \label{eqn:tot_power1}\\\!\!\!\!\!\!\!\!\!&& \;\;\;\; \gamma_R  \geq \gamma_{R,th}. \label{eqn:radar_snr_cons1}
	\end{eqnarray}
\end{subequations}
\vspace{-5mm}

\noindent Note that $R_u$, $R_{te}$, and $\gamma_R$ are all second order functions of $\mathbf w$ and $\mathbf W_n$, as (\ref{eqn:R_u}), (\ref{eqn:R_te}) and (\ref{eqn:gamma_R}), respectively. However, both  $R_u$ and $R_{te}$ are fractional forms of the design variables. Here, we invoke the quadratic transform technique \cite{Shen2018fractional} to recast $R_u$ and $R_{te}$ into non-fractional functions of $\mathbf w$ and $\mathbf W_n$.  Thereby, the problem $(\mathbb P_{1})$ is modeled as a non-fractional quadratic programming problem. The Lagrangian dual expressions of $R_u$ and $R_{te}$ in (\ref{eqn:R_u}) and (\ref{eqn:R_te}) are respectively
\begin{eqnarray}
	 \!\!\!\!\!\!\!\!\!\!\!&&R_u 
	  =  {\left( 1+\gamma_{u} \right)| \mathbf c_u^T \mathbf w|^2 }/\left({ | \mathbf c_u^T \mathbf w|^2 + || \mathbf c_u^T \mathbf W_n||^2 + \sigma_u^2 }\right) - \gamma_{u}  \nonumber \\ \!\!\!\!\!\!\!\!\!\!\!\!\!\!\! && + \log \left( 1+\gamma_{u} \right) =  - \alpha_u^2 (| \mathbf c_u^T \mathbf w|^2+|| \mathbf c_u^T \mathbf W_n||^2 + \sigma_u^2) \nonumber \\ \!\!\!\!\!\!\!\!\!\!\!\!\!\!\!&& + 2 \alpha_u \sqrt{1+\gamma_u} |\mathbf c_u^T \mathbf w| + \log \left( 1+\gamma_{u} \right) - \gamma_{u},\\
	   \!\!\!\!\!\!\!\!\!\!\!\!\!\!\!&&R_{te} 
	  =  {\left( 1+\gamma_{te} \right)| \mathbf c_{te}^T \mathbf w|^2 } \!/\! \left({ | \mathbf c_{te}^T \mathbf w|^2 \!+\!|| \mathbf c_{te}^T \mathbf W_n||^2 \!+\! \sigma_{te}^2 }\right)  - \gamma_{te} \nonumber \\ \!\!\!\!\!\!\!\!\!\!\!\!\!\!\! && + \log \left( 1+\gamma_{te} \right) =  - \alpha_{te}^2 (| \mathbf c_{te}^T \mathbf w|^2 + || \mathbf c_{te}^T \mathbf W_n||^2 + \sigma_{te}^2) \nonumber \\ \!\!\!\!\!\!\!\!\!\!\!\!\!\!\!&&+2 \alpha_{te} \sqrt{1+\gamma_{te}} |\mathbf c_{te}^T \mathbf w|  + \log \left( 1+\gamma_{te} \right) - \gamma_{te},
\end{eqnarray}
 \noindent where $\gamma_{u}$, $\alpha_u$,  $\gamma_{te}$, and $\alpha_{te}$ are auxiliary variables, whose values are updated in each iteration. Then, the objective can be re-written as 
 \begin{eqnarray}
 	R_u - R_{te} = c + \Re{(\mathbf v^T \mathbf w)} + \tr{\mathbf M (\mathbf W_n \mathbf W_n^H+\mathbf w \mathbf w^H)},
 \end{eqnarray}
\noindent where 
\begin{eqnarray}
	\!\!\!\!\!\!\!\!\!\!\!\!\!\!\!\!\!\!\!\!\!&&c = \log \left( 1\!+\!\gamma_{u} \right) \!-\! \gamma_u \!-\! \log \left( 1\!+\!\gamma_{te} \right) \!+\!\gamma_{te}\! +\! \alpha_{te}^2 \sigma_{te}^2 \!-\! \alpha_{u}^2 \sigma_{u}^2, \\\!\!\!\!\!\!\!\!\!\!\!\!\!\!\!\!\!\!\!\!\!&&
	\mathbf v = 2 \alpha_u \sqrt{1\!+\!\gamma_u} \mathbf c_u - 2 \alpha_{te} \sqrt{1\!+\!\gamma_{te}} \mathbf c_{te}, \label{eqn:v_vector} \\\!\!\!\!\!\!\!\!\!\!\!\!\!\!\!\!\!\!\!\!\!&& \mathbf M = \alpha_{te}^2  \mathbf c_{te}^{\ast} \mathbf c_{te}^{T} - \alpha_{u}^2  \mathbf c_{u}^{\ast} \mathbf c_{u}^{T}. \label{eqn:M_matrix}
\end{eqnarray}
 
\noindent Thereby, a quadratic programming problem with non-fractional objective and constraints is formulated. We invoke semidefinite relaxation (SDR) to address the re-formulated problem. By letting $\mathbf R_w =  \mathbf w^H \mathbf w$ and $\mathbf R_{W_n} =  \mathbf W_n^H \mathbf W_n$, the Problem  $(\mathbb P_{1})$ in (\ref{eqn:problem1}) becomes
\begin{subequations} \label{eqn:problem1_new}
	\begin{eqnarray} 
		\!\!\!\!\!\!\!\!\!\!\!(\overline{\mathbb P}_1) \!\!\!\!	\quad \max_{\mathbf w, \mathbf R_w \mathbf R_{W_n} }&&\!\!\!\!\!\!\!\! \Re{(\mathbf v^T \mathbf w)} + \tr{\mathbf M (\mathbf R_{W_n} + \mathbf R_{w})} + c  \label{eqn:obj1_new}\\ \!\!\!\!\!\!\!\!\!\!\!\mathrm{s.t.} && \!\!\!\!\!\!\!\! \tr{\mathbf R_w + \mathbf R_{W_n}}  \leq P_R \label{eqn:tot_power1_new}\\\!\!\!\!\!\!\!\!\!\!\!&& \!\!\!\!\!\!\!\! \tr{ \mathbf C_{T}^H \mathbf C_{T} \left[ \mathbf R_{w} + \mathbf R_{W_n} \right]  }/ \sigma_R^2  \geq \gamma_{R,th} \label{eqn:radar_snr_cons1_new} \\\!\!\!\!\!\!\!\!\!\!\!&& \!\!\!\!\!\!\!\! \mathbf R_w \succcurlyeq  \mathbf w \mathbf w^H
	\end{eqnarray}
\end{subequations}
\noindent where $c$ in the objective (\ref{eqn:obj1_new}) is irrelevant to the design variables, and thus can be dropped; $\mathbf R_w$ and $\mathbf R_{W_n}$ are set as positive semidefinite matrices in the optimization process. The Problem $(\overline{\mathbb P}_1)$ in (\ref{eqn:problem1_new}) is a linear programming problem, of which the optimal solution, say $\mathbf w^{\ast}$ and $\mathbf R_{W_n}^{\ast}$, can be obtained by numerical solvers, for example CVX toolbox \cite{cvx}. The optimal solution for $\mathbf W_n$ is obtained by calculating the square root matrix of $\mathbf R_{W_n}^{\ast}$.

\subsection{Second sub-problem: Solve for $\mathbf \Phi$ by fixing  $\mathbf w$ and $\mathbf W_n$}\label{sec:problem2}
The design problem of IRS parameter, $\mathbf \Phi$,  is as follows

\vspace{-5mm}
\begin{subequations} \label{eqn:problem2}
	\begin{eqnarray} 
	({\mathbb P}_2) \qquad	\max_{\mathbf \Phi}&&	\;\;\;\; R_u - R_{te}	\label{eqn:obj2}\\ \mathrm{s.t.} && \;\;\;\; 
		|\mathbf \Phi_{n,n}| = 1,\;\; \forall n = 1, \cdots, N\label{eqn:unit_modu2} \\ && \;\;\;\; \gamma_R \geq \gamma_{R,th} \label{eqn:radar_snr2}
	\end{eqnarray}
\end{subequations}
Besides the fractional form objective, the problem $({\mathbb P}_2)$ in (\ref{eqn:problem2}) is subject to challenging non-convex UMC on $\mathbf \Phi$ as (\ref{eqn:unit_modu2}). Moreover, the $\gamma_R$ in (\ref{eqn:radar_snr2}) is a non-convex fourth order function of $\mathbf \Phi$, and this term needs to be convexified to make the problem solvable. 

We re-write the effective channel for the user as
\begin{eqnarray}
	\!\!\!\!\!\!\!\!&&\mathbf c_u^T = 	\mathbf g^T + \beta_H^{\frac{1}{2}} \mathbf f^T \mathbf \Phi \mathbf H_{dl} =  \mathbf g^T + \boldsymbol{\phi}^T \beta_H^{\frac{1}{2}} \text{diag} (\mathbf f)  \mathbf H_{dl} \nonumber\\ \!\!\!\!\!\!\!\!&&=\mathbf g^T + \boldsymbol{\phi}^T \mathbf D, \label{eqn:c_u_T}
\end{eqnarray}
\noindent where $\boldsymbol{\phi} = \text{diag}(\mathbf \Phi)$ is the column vector that contains all diagonal elements of $\mathbf \Phi$, and $\mathbf D = \beta_H^{\frac{1}{2}} \text{diag} (\mathbf f)  \mathbf H_{dl}$. Similarly, the effective channel for the ED/target is re-written as
\begin{eqnarray}
	\!\!\!\!\!\!\!\!&&\mathbf c_{te}^T = \beta^{\frac{1}{2}} \mathbf a_I^T(\psi_{a},\psi_{e}) \mathbf \Phi \mathbf H_{dl} = \boldsymbol{\phi}^T \beta^{\frac{1}{2}} \text{diag} (\mathbf a_I(\psi_{a},\psi_{e}))  \mathbf H_{dl}\nonumber \\ \!\!\!\!\!\!\!\!&&= \boldsymbol{\phi}^T \mathbf E, \label{eqn:c_te_T}
\end{eqnarray}
\noindent where $\mathbf E=\beta^{\frac{1}{2}} \text{diag} (\mathbf a_I(\psi_{a},\psi_{e}))  \mathbf H_{dl}$. By invoking (\ref{eqn:v_vector}), (\ref{eqn:c_u_T}), and (\ref{eqn:c_te_T}), the first term in the transformed non-fractional objective in (\ref{eqn:problem1_new}), $\Re{(\mathbf v^T \mathbf w)}$, can be re-written as 
\begin{eqnarray}
	\Re{(\mathbf v^T \mathbf w)} = \Re{(\boldsymbol{\phi}^T \boldsymbol \eta)} + c_1,
\end{eqnarray}
\noindent where 
\begin{eqnarray}
\boldsymbol{\eta}&=&2(\alpha_u \sqrt{1+\gamma_u} \mathbf D - \alpha_{te} \sqrt{1+\gamma_{te}} \mathbf E) \mathbf w,\\ c_1 &=& 2 \Re{(\alpha_u \sqrt{1+\gamma_u}} \mathbf g^T \mathbf w).
\end{eqnarray}
Furthermore, $c_1$ is taken as a constant in the second sub-problem. Likewise, by referring to (\ref{eqn:M_matrix}),  (\ref{eqn:c_u_T}), and (\ref{eqn:c_te_T}),  the second item of the converted objective in (\ref{eqn:problem1_new}), $\tr{\mathbf M \mathbf R_{W_n}}$, is re-expressed as
\begin{eqnarray}
    &&\tr{\mathbf M \mathbf R_{W_n}} = \alpha_{te}^2  \mathbf c_{te}^{T} \mathbf W_n \mathbf W_n^H \mathbf c_{te}^{\ast} - \alpha_{u}^2  \mathbf c_{u}^{T} \mathbf W_n \mathbf W_n^H \mathbf c_{u}^{\ast} \nonumber \\ &&= \boldsymbol \phi^T \mathbf L_1 \boldsymbol \phi^{\ast} -\Re{(\boldsymbol{\phi}^T \boldsymbol \mu)} - c_2,
\end{eqnarray}
\noindent where 
\begin{eqnarray}
    \mathbf L_1 &=& \alpha_{te}^2  \mathbf E \mathbf W_n \mathbf W_n^H \mathbf E^H - \alpha_{u}^2  \mathbf D \mathbf W_n \mathbf W_n^H \mathbf D^H,\\ \boldsymbol \mu &=&2 \alpha_{u}^2  \mathbf D \mathbf W_n \mathbf W_n^H \mathbf g^{\ast}, \\ c_2 &=& \alpha_{u}^2  \mathbf g^T \mathbf W_n \mathbf W_n^H \mathbf g^{\ast}.
\end{eqnarray}
Likewise, $c_2$ is taken as a constant in the design of $\mathbf \Phi$ or 
$\boldsymbol{\phi}$. Similarly, $\tr{\mathbf M \mathbf R_{w}}$ can be re-written as
\begin{eqnarray}
    &&\tr{\mathbf M \mathbf R_{w}} = \boldsymbol \phi^T \overline{\mathbf L}_1 \boldsymbol \phi^{\ast} -\Re{(\boldsymbol{{\phi}^T \overline{\boldsymbol \mu})}} - \bar c_2,
\end{eqnarray}
\noindent where 
\begin{eqnarray}
    \overline{\mathbf L}_1 &=& \alpha_{te}^2  \mathbf E \mathbf w \mathbf w^H \mathbf E^H - \alpha_{u}^2  \mathbf D \mathbf w \mathbf w^H \mathbf D^H,\\ \overline{\boldsymbol \mu} &=&2 \alpha_{u}^2  \mathbf D \mathbf w \mathbf w^H \mathbf g^{\ast}, \\ \bar c_2 &=& \alpha_{u}^2  \mathbf g^T \mathbf w \mathbf w^H \mathbf g^{\ast}.
\end{eqnarray}
Thereby, the objective, $R_u - R_{te}$ can be re-written as
\begin{eqnarray}
\!\!\!\!\!\!\!\!\!\!\!\!	&&R_u - R_{te} = \Re{(\mathbf v^T \mathbf w)} + \tr{\mathbf M (\mathbf R_{W_n} + \mathbf R_{w})} + c \nonumber \\\!\!\!\!\!\!\!\!\!\!\!\!&&= \boldsymbol \phi^T (\mathbf L_1 +\overline{\mathbf L}_1)\boldsymbol \phi^{\ast} +\Re{[\boldsymbol{\phi}^T (\boldsymbol \eta-\boldsymbol \mu - \overline{\boldsymbol \mu})]} + \text{const}, \label{eqn:obj2_new}
\end{eqnarray}
\noindent which is a quadratic function of $\boldsymbol{\phi}$.

Next, we need to transform the radar SNR, $\gamma_R$, in (\ref{eqn:radar_snr2}) to make $({\mathbb P}_2)$ solvable. Recall that the radar channel $\mathbf C_T = \beta \mathbf H_{ul} \mathbf \Phi \mathbf a_I(\psi_{a},\psi_{e}) \mathbf a_I^{T}(\psi_{a},\psi_{e}) \mathbf \Phi \mathbf  H_{dl}$. By denoting $\mathbf U = \mathbf \Phi \mathbf a_I(\psi_{a},\psi_{e}) \mathbf a_I^{T}(\psi_{a},\psi_{e}) \mathbf \Phi$, $\gamma_R$ can be re-arranged as
\begin{eqnarray}
	&&\gamma_R= \tr{ \mathbf C_{T}^H \mathbf C_{T} \left[ \mathbf w \mathbf w^H + \mathbf W_n \mathbf W_n^H \right]  }/ \sigma_R^2 \nonumber \\&&= \beta^2/\sigma_R^2 \tr{  \mathbf U^H \mathbf  H_{ul}^H \mathbf  H_{ul} \mathbf  U \mathbf  H_{dl} (\mathbf w \mathbf w^H + \mathbf W_n \mathbf W_n^H) \mathbf  H_{dl}^H} \nonumber \\&&\overset{(a)}{=} \beta^2/\sigma_R^2 \mathbf u^H \mathbf Z \mathbf u, \label{eqn:gamma_R_new}
\end{eqnarray}
\noindent where in step (a), $\tr{  \mathbf U^H \mathbf A \mathbf U \mathbf B^T} = \mathbf u^H (\mathbf B \otimes \mathbf A) \mathbf u$ is referred to, $\mathbf Z = [\mathbf  H_{dl} (\mathbf w \mathbf w^H + \mathbf W_n \mathbf W_n^H) \mathbf  H_{dl}^H]^T \otimes (\mathbf  H_{ul}^H \mathbf  H_{ul})$, and $\mathbf u = \text{vec}(\mathbf U)$. Note that both $\mathbf U$ and $\mathbf u$ are quadratic in $\mathbf \Phi$ or $\boldsymbol{\phi}$. To create a lower bound for $\gamma_R$ that is quadratic in $\boldsymbol{\phi}$, we need to havea bound which is linear in $\mathbf u$. For this we invoke MM \cite{Sun2017majorization} as
\begin{eqnarray}
	&&\gamma_R = \beta^2/\sigma_R^2 \mathbf u^H \mathbf Z \mathbf u \nonumber \\ &&\geq \tilde \gamma_R = \beta^2/\sigma_R^2 (\mathbf u^H \mathbf Z \mathbf u_t + \mathbf u_t^H \mathbf Z \mathbf u - \mathbf u_t^H \mathbf Z \mathbf u_t), \label{eqn:tilde_gamma_R}
\end{eqnarray}
\noindent where $\mathbf u_t = \text{vec}[\mathbf \Phi_t \mathbf a_I(\psi_{a},\psi_{e}) \mathbf a_I^{T}(\psi_{a},\psi_{e}) \mathbf \Phi_t]$, where $\mathbf \Phi_t$ is the solution of $\mathbf \Phi$ in $t$-th/previous iteration,  the current iteration index is $(t+1)$, and $\tilde \gamma_R$ is the surrogate function for $\gamma_R$. In addition, the first term of $\tilde \gamma_R$ is re-expressed as
\begin{eqnarray}
	&&\mathbf u^H \mathbf Z \mathbf u_t \overset{(b)}{=} \text{vec}(\mathbf U)^H \text{vec}(\mathbf V) = \tr{\mathbf U^H \mathbf V} \nonumber \\&& \overset{(c)}{=} \tr{\mathbf \Phi^H \mathbf a_I^{\ast}(\psi_{a},\psi_{e}) \mathbf a_I^{H}(\psi_{a},\psi_{e}) \mathbf \Phi^H \mathbf V} \nonumber \\&&= \boldsymbol \phi^H \{[\mathbf a_I^{\ast}(\psi_{a},\psi_{e}) \mathbf a_I^{H}(\psi_{a},\psi_{e})] \circ \mathbf V^T\} \boldsymbol \phi^{\ast},
\end{eqnarray}
\noindent where  $\mathbf u = \text{vec}(\mathbf U)$; $\mathbf V$ is set such that $\text{vec}(\mathbf V) = \mathbf Z \mathbf u_t$;  $\mathbf U = \mathbf \Phi \mathbf a_I(\psi_{a},\psi_{e}) \mathbf a_I^{T}(\psi_{a},\psi_{e}) \mathbf \Phi$. Similarly, the second term in $\tilde \gamma_R$, $\mathbf u_t^H \mathbf Z \mathbf u$, can be re-written as
\begin{eqnarray}
	\mathbf u_t^H \mathbf Z \mathbf u = \boldsymbol \phi^T \{[\mathbf a_I(\psi_{a},\psi_{e}) \mathbf a_I^{T}(\psi_{a},\psi_{e})] \circ \mathbf Y^T\} \boldsymbol \phi,
\end{eqnarray}
\nonumber where the matrix $\mathbf Y$ satisfies $\text{vec}(\mathbf Y) = \mathbf Z^T \mathbf u_t^{\ast}$. Thereby, $\tilde \gamma_R$ in (\ref{eqn:tilde_gamma_R}) can be written as
\begin{eqnarray}
	\tilde \gamma_R = \boldsymbol \phi^H \mathbf L_2 \boldsymbol \phi^{\ast} + \boldsymbol \phi^T \mathbf L_3 \boldsymbol \phi -\beta^2/\sigma_R^2   \mathbf u_t^H \mathbf Z \mathbf u_t, \label{eqn:tilde_gamma_R_new}
\end{eqnarray}
\noindent where the third term is not relevant to the variable $\boldsymbol \phi$, and
\begin{eqnarray}
	&&\mathbf L_2 = \beta^2/\sigma_R^2 \{[\mathbf a_I^{\ast}(\psi_{a},\psi_{e}) \mathbf a_I^{H}(\psi_{a},\psi_{e})] \circ \mathbf V^T\},  \\ &&\mathbf L_3 = \beta^2/\sigma_R^2 \{[\mathbf a_I(\psi_{a},\psi_{e}) \mathbf a_I^{T}(\psi_{a},\psi_{e})] \circ \mathbf Y^T\}.
\end{eqnarray}
 
So far, the objective (\ref{eqn:obj2}) and constraint (\ref{eqn:radar_snr2}) in $	({\mathbb P}_2)$ are transformed into non-fractional quadratic form. By invoking (\ref{eqn:obj2_new}) and (\ref{eqn:tilde_gamma_R_new}), $	({\mathbb P}_2)$ can be re-written as

\begin{subequations} \label{eqn:problem2_newer}
	\begin{eqnarray} 
	\!\!\!\!\!\!\!\!\!\!	(\overline{\mathbb P}_2) \quad	\max_{ \boldsymbol{\phi}}&&	\!\!\!\!\!\boldsymbol \phi^T (\mathbf L_1 +\overline{\mathbf L}_1)\boldsymbol \phi^{\ast} +\Re{[\boldsymbol{\phi}^T (\boldsymbol \eta-\boldsymbol \mu - \overline{\boldsymbol \mu})]}	\label{eqn:obj2_newer}\\ \mathrm{s.t.} && \!\!\!\!\! 
		|\boldsymbol \phi_{n,1}| = 1,\;\; \forall n = 1, \cdots, N\label{eqn:unit_modu2_newer} \\ && \!\!\!\!\! \boldsymbol \phi^H \mathbf L_2 \boldsymbol \phi^{\ast} + \boldsymbol \phi^T \mathbf L_3 \boldsymbol \phi\geq  \gamma_{R,th}' \label{eqn:radar_snr2_newer}
	\end{eqnarray}
\end{subequations}
\noindent  where $\gamma_{R,th}' = \gamma_{R,th} + \beta^2/\sigma_R^2   \mathbf u_t^H \mathbf Z \mathbf u_t $. Now, $(\overline{\mathbb P}_2)$ is a quadratic programming problem with UMC on the elements of the variable $\boldsymbol \phi$ as (\ref{eqn:unit_modu2_newer}). We solve $(\overline{\mathbb P}_2)$ with SDR {by relaxing the UMC as }
\begin{subequations} \label{eqn:problem2_newest}
	\begin{eqnarray} 
		\!\!\!\!\!\!\!\!\!\!	(\overline{\overline{\mathbb P}}_2) 	\max_{\mathbf R_1, \mathbf R_2, \boldsymbol{\phi}}&&	\!\!\! \!\!\! \tr{( \mathbf L_1 +\overline{\mathbf L}_1)\mathbf R_1^{\ast}} +\Re{[\boldsymbol{\phi}^T (\boldsymbol \eta-\boldsymbol \mu - \overline{\boldsymbol \mu})]}	\label{eqn:obj2_newest}\\ \mathrm{s.t.} && \!\!\!\!\!\!
		[\mathbf R_1]_{n,n} = 1, \forall n = 1, \cdots, N \label{eqn:R1} \\&&  \!\!\!\!\!\!
		|[\mathbf R_2]_{n,n}| \leq 1, \forall n = 1, \cdots, N \label{eqn:R2} \\ &&  \!\!\!\!\!\!\tr{\mathbf L_2 \mathbf R_2^{\ast}}  + \tr{\mathbf L_3 \mathbf R_2}   \geq   \gamma_{R,th}' \label{eqn:radar_snr2_newest} \\ && \!\!\!\!\!\! \mathbf R_1 \succcurlyeq \boldsymbol{\phi} \boldsymbol{\phi}^H
		\\ && \!\!\!\!\!\! \mathbf R_2 \succcurlyeq \boldsymbol{\phi} \boldsymbol{\phi}^T
	\end{eqnarray}
\end{subequations} 
\noindent The solved $\boldsymbol{\phi}$ is substituted into the sub-problem 1 of next iteration. In addition, the overall optimization algorithm for jointly designing $\mathbf w$, $\mathbf W_n$ and  $\mathbf \Phi$, is summarized as Algorithm \ref{alg:alt_opt}. Furthermore,  $\mathrm{obj}^{(t+1)}$ is the objective, or secrecy rate $(R_u-R_{te})$, obtained in the $(t+1)$-th iteration. $\varepsilon$ is the indicator of error tolerance.

\begin{algorithm}\label{alg:alt_opt}
	\SetAlgoLined
	\KwResult{Return $\mathbf w$, $\mathbf W_n$ and  $\mathbf \Phi$.}
	\textbf{Initialization:} $\mathbf \Phi = \mathbf \Phi_0$, $\mathbf w = \mathbf w_0$, $\mathbf W_n = \mathbf W_{n,0}$, $\gamma_u =\gamma_{u,0}$, $\alpha_u =\alpha_{u,0}$, $\gamma_{te} =\gamma_{te,0}$, $\alpha_{te} =\alpha_{te,0}$,  $t=0$\;
	\While{ $\!\!\!(1)$ }{
		
		\textbf{ \emph{// Auxiliary variables update}}
		
		$\alpha_u = {\sqrt{1\!+\!\gamma_u}| \mathbf c_u^T \mathbf w| }/{\left( | \mathbf c_u^T \mathbf w|^2 \!+\! || \mathbf c_u^T \mathbf W_n||^2 \!+\! \sigma_u^2 \right)}$\;
		
		$\gamma_u = {| \mathbf c_u^T \mathbf w|^2 }/{\left( || \mathbf c_u^T \mathbf W_n||^2 + \sigma_u^2 \right)}$\;
		
		$\alpha_{te} = {\sqrt{1 \!+\!\gamma_{te}}| \mathbf c_{te}^T \mathbf w| }/{\left( | \mathbf c_{te}^T \mathbf w|^2 \!\!+\! || \mathbf c_{te}^T \mathbf W_n||^2 \!\!+ \!\sigma_{te}^2 \right)}$\;
		
		$\gamma_{te} = {| \mathbf c_{te}^T \mathbf w|^2 }/{\left( || \mathbf c_{te}^T \mathbf W_n||^2 + \sigma_{te}^2 \right)}$\;

		\textbf{ \emph{// First sub-problem} }
		
		Solve $(\overline{\mathbb P}_1)$ of (\ref{eqn:problem1_new}) to obtain $\mathbf w$ and $\mathbf R_{W_n}$\;

  Obtain $\mathbf W_n$ as the 
		square root matrix of $\mathbf R_{W_n}$ \;

		\textbf{  \emph{//  Second sub-problem} }
		
		Substitute $\mathbf w$, $\mathbf W_n$, found  in the first sub-problem  into the second sub-problem\;
		
		Obtain $\boldsymbol{\phi}$ by solving $(\overline{\overline{\mathbb P}}_2)$ in (\ref{eqn:problem2_newest})\;
		
		$\mathbf \Phi = \text{diag}(\boldsymbol\phi)$\;

		\textbf{ \emph{// Condition of termination}}
		
		\If{$\left((t\geq t_{\max})\;||\;(\big|\!\left[ \mathrm{obj}^{(t+1)} \!\!-\! \mathrm{obj}^{(t)}\right]\!/\mathrm{obj}^{(t)}\!\big| \leq \varepsilon)\right)$}{
			\texttt{Break}\;
		}
		
		$t = t+1$;
		
	}
	\caption{Secrecy rate maximization}
\end{algorithm}
\vspace{-0mm}

\begin{table}[]
	\begin{center}			
		\caption{System Configurations}
		\label{tab:sys_para}
		\begin{tabular}{p{6.5cm}|p{1.0cm}} 
			\hline
			\textbf{Parameter} & \textbf{Value}
			\\\hline
			Error tolerance indicator of Algorithm \ref{alg:alt_opt},  $\varepsilon$ [dB] & $- 20$
			\\\hline
			Maximum number of iterations allowed, $t_{\max}$ & $20$			
			\\\hline
			Rician channels $\mathbf g$, $\mathbf f$, $\mathbf H_{dl}$, and $\mathbf H_{ul}$ [dB] & $0$
			\\\hline
			  Noise power at the radar receiver,  $\sigma_R^2$ [dBm]&$0$
			\\\hline
			Noise power at the communication user, $\sigma_u^2$ [dBm]&$0$
			\\\hline
			Noise power at the target/ED, $\sigma_{te}^2$ [dBm]&$0$
			\\\hline
			  Inter-antenna distance at radar transmitter/receiver, $d$ & $ \lambda/2$
			\\\hline
			Number of  antennas of radar transmitter, $N_T$ & $16$
			\\\hline
			Number of  antennas of radar receiver, $N_R$ & $16$
			\\\hline
			Number of  IRS elements, $N$ & $25$
			\\\hline
			Target coefficient $|\beta|$ [dB] & $-40$
			\\\hline
			SNR threshold at radar receiver $\gamma_{R,th}$ [dB] & $-11$
			\\\hline
		Radar power budget $P_R$ [dBm] & $30$
			\\\hline

		\end{tabular}
	\end{center}
	\vspace{-0mm}
\end{table}

\section{NUMERICAL RESULTS}
In this section, we present numerical results to demonstrate the convergence of the proposed algorithm. 
{While AN helps PLS, it results in  loss of communication signal power. Let $\omega$ be ratio of 
user information power  over the total power of user information and AN.  
We    show via simulations  of $\omega$, on the system performance}. The channels are simulated as Rician. The parameters of the simulation are shown as in Table \ref{tab:sys_para}.

\begin{figure}[!t]\centering\vspace{-5mm}
	\includegraphics[width=0.42\textwidth]{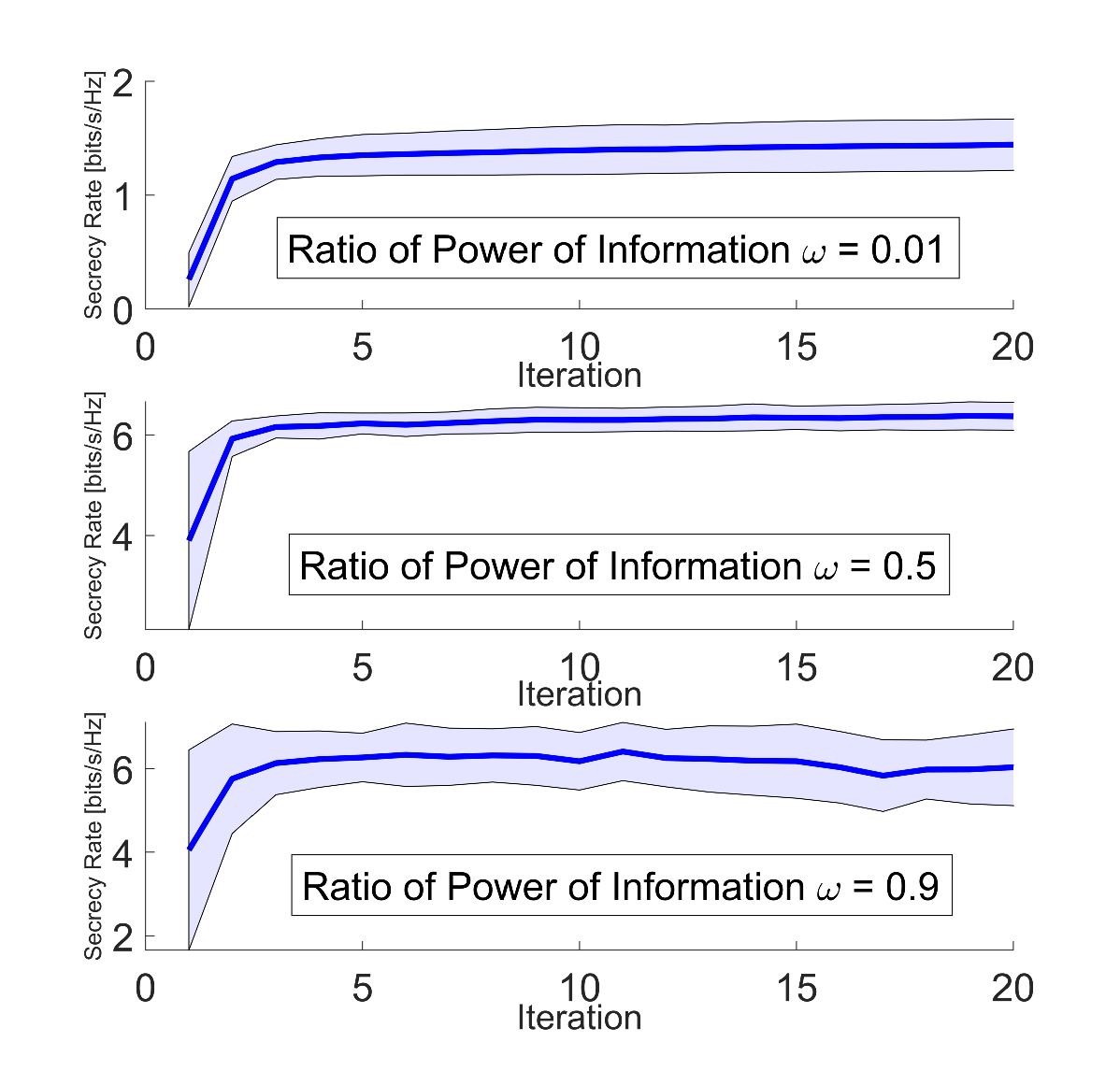}
	\caption{Convergence of the proposed algorithm.}
	\label{fig:Fig_convergence_sr}\vspace{-3mm}
\end{figure}

Fig. \ref{fig:Fig_convergence_sr} demonstrates the convergence of our proposed algorithm. The solid blue line represents the mean of the convergence curves of the objective (secrecy rate), which are obtained with $30$ different channel realizations. The light blue shaded area around the solid line shows the variance among the $30$ distinct realizations. It is observed that, the objective, $R_u - R_{te}$,  reaches convergence in few iterations. In addition, the convergence holds for different values of $\omega$. 

\begin{figure}[!t]\centering\vspace{-0mm}
	\includegraphics[width=0.38\textwidth]{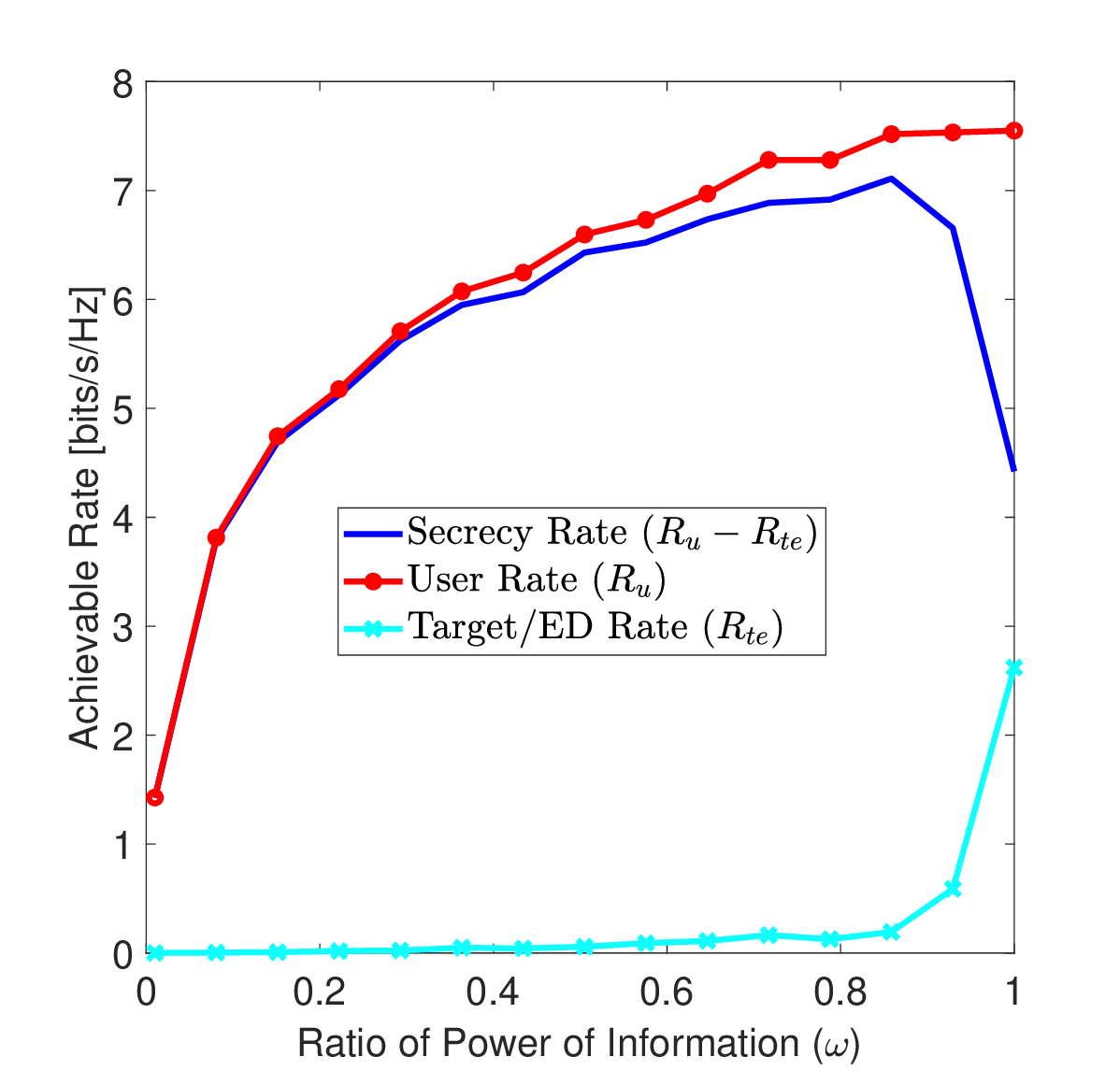}
	\caption{The impact of $\omega$ on the achievable rates.}
	\label{fig:Fig_rate_vs_ratio}\vspace{-3mm}
\end{figure}

{
In Fig. \ref{fig:Fig_rate_vs_ratio}, the influence of the power ratio of user information, $\omega$, on the achievable rates, is investigated. When $\omega$ is small, low power is allocated to the user information, and high power to the AN. As a result, the user rate ($R_u$) and rate at the target/ED ($R_{te}$) are both low, and the secrecy rate ($R_u-R_{te}$) is also low in this case. When $\omega$ increases, both $R_u$ and $R_{te}$ are enhanced. $R_u$ increases fast when $\omega$ is small, and  slow when $\omega$ is large. Meanwhile, $R_{te}$ rises slowly when $\omega$ is low, and increases fast when $\omega$ is large. Therefore, with the increase of $\omega$, the secrecy rate, $R_u-R_{te}$, increases at first, since $R_u$ increases faster than $R_{te}$ in the beginning (See Fig. \ref{fig:Fig_rate_vs_ratio} when $\omega$ is less than  $0.86$). Afterwards, $R_{te}$ rises faster, so the secrecy rate drops. When $\omega$ is $1$, only user information embedded waveform is transmitted towards and illuminate the target/ED for the sensing task, which leads to  high ED rate, and decreased secrecy rate.}

\vspace{2mm}
\section{CONCLUSIONS}
\label{sec:conclusions}
In this paper, we have considered an IRS-aided DFRC system, where the target to be sensed acts as an eavesdropper. We have proposed an alternating optimization algorithm to jointly design the transmitted  waveform for user information, the AN, and the IRS parameter matrix, to maximize the secrecy rate subject to certain non-convex constraints. We have invoked an efficient fractional programming technique, quadratic transform, to convert the fractional objective of secrecy rate, to a more mathematically tractable non-fractional form. Thereby, the design of  waveform of user information and AN is transformed to a non-fractional quadratic programming problem. The IRS design problem contains a challenging fourth order function of the IRS parameter, which is degraded to a  second order one via the effective bounding technique, MM, to deliver a tractable IRS optimization problem. The numerical results have demonstrated  the convergence of the proposed algorithm, and the influence of the user information power ratio on the achievable rates.



\balance


\vspace{12pt}

\bibliographystyle{IEEEtran}
\bibliography{IEEEabrv,References,Ref2}

\end{document}